\documentclass[twoside,twocolumn,english,pra,aps,superscriptaddress]{revtex4-2}
\usepackage[T1]{fontenc}
\usepackage{color}
\usepackage{babel}
\usepackage{textcomp}
\usepackage{amsmath}
\usepackage{amssymb,amsfonts,mathrsfs}
\usepackage{amsthm}
\usepackage{graphicx}
\usepackage{float}
\usepackage[unicode=true,pdfusetitle,
bookmarks=true,bookmarksnumbered=false,bookmarksopen=false,
breaklinks=false,pdfborder={0 0 0},backref=false,colorlinks=true]
{hyperref}
\usepackage{breakurl}
\usepackage{listings}
\makeatletter
\theoremstyle{plain}

\makeatother

\begin{document}
	
\preprint{This line only printed with preprint option}
	
\title{Emergent Non-Abelian Thouless Pumping Induced by the Quasiperiodic Disorder}
	
\author{Sen Huang}
\affiliation{Guangdong Provincial Key Laboratory of Quantum Engineering and Quantum Materials, School of Physics and Telecommunication Engineering, South China Normal University, Guangzhou 510006, China}
	
\author{Yan-Qing Zhu}
\email{Corresponding author: yqzhuphy@hku.hk}
\affiliation{Department of Physics and HK Institute of Quantum Science and Technology, The University of Hong Kong, Pokfulam Road, Hong Kong, China}
\affiliation{Guangdong-Hong Kong Joint Laboratory of Quantum Matter, Frontier Research Institute for Physics, South China Normal University, Guangzhou 510006, China}

\author{Zhi Li}
\email{Corresponding author: lizphys@m.scnu.edu.cn}
\affiliation{Guangdong Provincial Key Laboratory of Quantum Engineering and Quantum Materials, School of Physics and Telecommunication Engineering, South China Normal University, Guangzhou 510006, China}
\affiliation{Guangdong-Hong Kong Joint Laboratory of Quantum Matter, Frontier Research Institute for Physics, South China Normal University, Guangzhou 510006, China}
\affiliation{Guangdong Provincial Key Laboratory of Nuclear Science, Institute of Quantum Matter, South China Normal University, Guangzhou 510006, China}

\date{\today}
	
\begin{abstract}
We investigate the non-Abelian Thouless pumping in a disorder tunable Lieb chain with degenerate flat bands. The results reveal that quasiperiodic disorder will cause a topological phase transition from the trivial (without non-Abelian Thouless pumping) to the non-trivial (with non-Abelian Thouless pumping) phase. The mechanism behind is that the monopole originally outside the topological region can be driven into the topological region due to the introduction of quasiperiodic disorder. Moreover, since the corresponding monopole will turn into a nodal line to spread beyond the boundaries of the topological region, the system with large disorder strength will result in the disappearance of non-Abelian Thouless pumping. Furthermore, we numerically simulate the Thouless pumping of non-Abelian systems, and the evolution results of center of mass’ displacement are consistent with the Chern number. Finally, we discuss the localization properties of the system and find that, similar to [PRL {\bf 130}, 206401(2023)], the inverse Anderson transition does not occur in the system with the increase of quasiperiodic strength, while the system still maintains the coexistence of localized and extended states.
\end{abstract}
	
\maketitle

\section{Introduction}
Thouless pumping as the quantized transport of particles has attracted intensive interest since D. J. Thouless proposed it in 1983~\cite{DJThouless1983}. In Thouless pumping, the transport of charge is related to the Chern numbers and shows the topological equivalence to the integer quantum Hall effect in two dimensions~\cite{KvKlitzing1980, QNiu1984, QNiu1985, DXiao2010}. Thouless pumping have been realized using different platforms~\cite{XLQi2011,MZHasan2010,DWZhang2018,NRCooper2019,NGoldman2016,TOzawa2019,MLohse2016,SNakajima2016,HILu2016,CSchweizer2016,SNakajima2021,AFabre2022,JMinguzzi2022,LWang2013,YEKraus2012,ACerjan2020,QCheng2022,WLiu2022,YKe2016,Wcheng2020,HILu2016} and the extension of the Thouless pumping includes spin pumping~\cite{CSchweizer2016,IHGrinberg2020}, nonlinear Thouless pumping~\cite{Mjurgensen2021,Mjurgensen2022,QFu2022,NMostaan2022,MJurgensen2023}, interacting topological pumping~\cite{RGawatz2022,JAMarks2021,YKe2017,TSZeng2015,ASWalter,YKe2023}, high-order topological pumping~\cite{WABenalcazar2022,JFWienand2022,BLWu2022}, and non-Hermitian topological pumping~\cite{CYuce2019,WHu2017,ZFedorova2020}. Recently, Thouless pumping has been extended to non-Abelian version~\cite{RCitro2023,VBrosco2021,OYou2022,YKSun2022,YYang2023,MParto2023}. With synthetic non-Abelian gauge fields~\cite{PHauke2012,LLepori2016,YYang2019,QXLv2023,DCheng2023,GPalumbo2021}, non-Abelian version of Thouless pumping has been constructed in theory~\cite{VBrosco2021} and realized in acoustic and photonic waveguides~\cite{OYou2022,YKSun2022}. In non-Abelian Thouless pumping, the quantization of the pumping is related to the Wilczek and Zee holonomy which is the non-Abelian analog of Berry's phase~\cite{FWilczek1984,DWZhang2020,SLZhuHFu2006,LBShao2008,SLZhu2007,SLZhu2006,XShen2018,XShen2019,XDHu2021,ZXGuo2022,XShen2022,HTDing2023}.

The competition between topology and disorder is another important issue~\cite{JQin,MMWauters2019,ALCHayward2021,JLi2009,CWGroth2009,CWGroth2009,CZChen2015,HMGuo2010,AAltland2014}. Thouless pumping is robust to disorders as long as the energy gap keeps opening~\cite{XLQi2011,MZHasan2010,DWZhang2018,SNakajima2021,ACerjan2020}. A strange disorder-induced topological phase called topological Anderson insulators(TAIs)~\cite{JLi2009,CWGroth2009} has been discovered and experimentally observed\cite{EJMeier,SStutzer2018,GGLiu2020}. In recent years, TAIs have been extended to various models, such as quasiperiodic SSH chains, $Z_2$ TAIs, long range SSH model, and non-Hermitian disordered systems\cite{LZTang2022, XHCui2022, HCHsu2020, LZTang2020}. Recently, disorder-induced quantized topological pumping have been demonstrated in noninteracting and interacting systems with different disorders which declares the existence of the topological Anderson Thouless pump~\cite{YPWu2022}. Based on the above previous studies, one can easily think of extending the study of the interplay between topology and disorder to cases involving non-Abelian gauge fields.

Here we explore the interplay between non-Abelian Thouless pumping and quasiperiodic disorder. Remarkably, a disorder-induced quantized non-Abelian topological pumping is proposed in noninteracting systems with quasiperiodic disorders. First we show the robustness of non-Abelian Thouless pumping under disorders, where topological pumping breaks for large disorder strength. Then we reveal a quantized non-Abelian topological pumping induced by quasiperiodic disorders from a trivial pump in the clean regime and explain the mechanism of the disorder-induced non-Abelian Thouless pumping as a result of the shift of the monopole. We further demonstrate that the disorder-induced non-Abelian Thouless pumping can also be observed in optical systems, similar to the topological pumping of light in waveguide arrays~\cite{QCheng2022,Mjurgensen2021}. Eventually, we give the inverse participation ratio (IPR) calculation of the disorder-induced non-Abelian topological pumping system to demonstrate a strange non-Abelian inverse Anderson transition which shows huge differences from the Anderson transition~\cite{WZhang2023}.

The paper is organized as follows. In Sec.~\ref{s2} we describe the non-Abelian Lieb chain model with quasiperiodic disorders. Then we show the robustness of non-Abelian Thouless pumping with quasiperiodic disorders which is similar to the Abelian case. We present the existence of disorder-induced non-Abelian Thouless pumping in Sec.~\ref{s3} and give a numerical simulation in optical waveguides to observe the topological pumping. Then we present numerical results of the IPR. Finally in Sec.~\ref{s4} we make conclusions.

\section{Model}
\label{s2}
\begin{figure}[bhtp] \centering
\includegraphics[width=8.5cm]{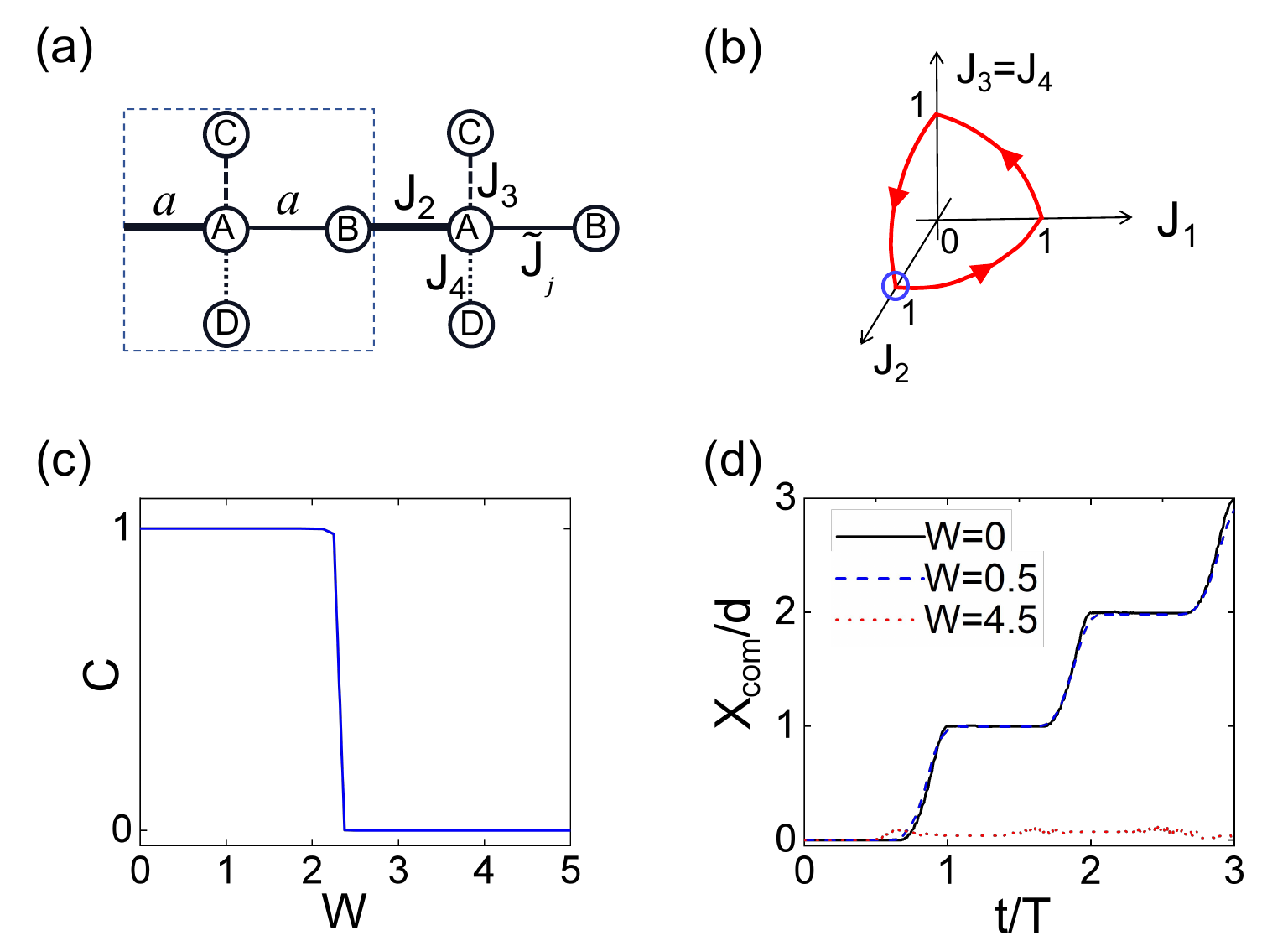}
\caption{(a) The non-Abelian quasiperiodic Lieb chain. The dashed box is the unit cell. The hopping strength and the lattice spacing $a$ are marked. (b) The corresponding adiabatic evolution loop, which reduces to the case without disorder for $W=0$. The red arrows indicate the pumping direction and the blue circle is the starting point. (c) Chern number $C$ as a function of $W$ for $L=10$ and $\beta=0$. (d) The corresponding time evolution of the center of mass $X_{com}\left(t \right)$ for the quasiperiodic disorder strengths $W=0$, $0.5$, $4.5$, respectively and $L=25$. The results are averaged over $50$ samples with $\beta$ valued from $0\sim 2\pi $ uniformly and $T=200$. The four hopping terms are evolving as shown in (b). Throughout, we set $a=1$.} \label{p1}
\end{figure}

Let's start at a one dimensional non-Abelian quasiperiodic Lieb chain model, which has four sites per unit cell as shown in Fig.~\ref{p1}(a)~\cite{VBrosco2021}. The corresponding Hamiltonian reads
\begin{equation}\label{ham}
  H=\sum_j{\left( \tilde{J}_{j}a_{j}^{\dagger}b_j+J_2a_{j}^{\dagger}b_{j-1}+J_3a_{j}^{\dagger}c_j+J_4a_{j}^{\dagger}d_j+H.c. \right)},
\end{equation}
where $a_{j}^{\dagger}(a_{j})$, $b_{j}^{\dagger}(b_{j})$, $c_{j}^{\dagger}(c_{j})$ and $d_{j}^{\dagger}(d_{j})$ denote the fermion creation (annihilation) operators at sites $A$, $B$, $C$ and $D$, respectively. The corresponding intercell hopping $\tilde{J}_{j}$ and intracell hopping $J_{2,3,4}$ are marked in Fig.~\ref{p1}. The disorder of the system is reflected in $\tilde{J}_{j}$ term, i.e., $\tilde{J}_{j}=J_1+W\cos\left( 2\pi\alpha j+\beta\right)$. $W$ is the disorder strength. $\alpha$ is an irrational number denoting the incommensurate modulation. Without loss of generality, we set $\alpha=\left(\sqrt{5}-1\right)/2$ in the following computation. $\beta$ is a phase shift. The Hamiltonian features two degenerate flat bands which construct a synthetic non-Abelian gauge field~\cite{VBrosco2021}. The corresponding energy spectrum is plotted in the Appendix~\ref{spectrum}. We consider noninteracting pumping in the lattice of size $4L$ with particle number $N=3L$, this is to say, we do a $3/4$ filling, which guarantee the Fermi energy is higher than the degenerate flat bands. 

One can use an adiabatic evolving loop to implement topological pumping in disorder tunable non-Abelian system [see red loop in Fig.~\ref{p1}(b)]. Similar methods have been used in dicussing non-Abelian physics without disorder~\cite{VBrosco2021}. To be more specific, the corresponding analytic expressions for different hopping terms have been provided in Appendix~\ref{AEHT}. For convenience, we set the distance between an abitrary point of the loop and the origin point equals to $1$, in other words, $\max[J_1,J_2,J_3,J_4]=1$ as shown in Fig.~\ref{p1}(b). To describe the topology of the system with Chern number under disorders, one can impose the twisted periodic boundary condition by introducing a twisted phase $\theta\in\left[0,2\pi\right]$ to the hopping terms. Then, one can obtain~\cite{QNiu1984,QNiu1985,DXiao2010,YPWu2022,llin2023}
\begin{equation}\label{hamtheta}
\begin{split}
	H_{\theta}=\sum_j{\left( \tilde{J}_je^{i\theta}a_{j}^{\dagger}b_j+J_2e^{i\theta}a_{j}^{\dagger}b_{j-1}+J_3e^{i\theta}a_{j}^{\dagger}c_j \right.}
	\\
	\left. +J_4e^{i\theta}a_{j}^{\dagger}d_j+H.c. \right).
\end{split}
\end{equation}
Note that, in the thermodynamic limit ($L\rightarrow\infty$), the Hamiltonian with a twisted phase $H_{\theta}$ is topologically equivalent to $H$ in Eq.~\eqref{ham}, so the two have the same Chern number.

We next discuss the relation between the non-Abelian Thouless pumping and the Chern number $C$ of $H_{\theta}$ under time-modulated parameters $\tilde{J}_j(t)$ and $J_{2,3,4}(t)$. As we know that the shift of the center-of-mass of Wannier state is guaranteed by the Chern number, i.e., $\Delta X_{com}=X_{com}(T)-X_{com}(0)=C$\cite{DWZhang2018,MLohse2016,SNakajima2016,HILu2016,SNakajima2021,ACerjan2020,YPWu2022}. Here the Wannier center for each unit cell in real space is associated with the Zak phase or polarization, i.e., $X_{com}(t)=\frac{1}{2\pi}\text{tr}~\gamma(t)$, where $\gamma$ denotes the non-Abelian Zak phase (or Berry-Wilczek-Zee phase \cite{FWilczek1984}) defined as
\begin{equation}\label{Zak}
  \gamma= \int_0^{2\pi} d\theta A_{\theta},
\end{equation}
with the non-Abelian Berry connection $A_{\theta}=\langle\psi|i\partial_{\theta}|\psi \rangle$ and its component $A_{\theta}^{nm}=\langle\psi_n |i\partial _{\theta}| \psi_m \rangle$. Here $|\psi_m \rangle$ indicates the eigenstate of the degenerate flat bands with the index ranging from $L+1$ to $3L$\cite{Bohm2003}. Note that, only the middle degenerate flat bands contribute to the non-trivial Chern number in the non-Abelian Thouless pumping and thus we mainly focus on the physics of these degenerate flat bands hereafter. One can also find that the non-Abelian Zak phase is a matrix instead of being a scalar in the Abelian case.

By modulating the paremeters in one loop period $T$, we have \cite{DWZhang2018, YPWu2022}
\begin{equation}
\begin{split}
\Delta X_{com}&=\frac{1}{2\pi}\text{tr}\left[\gamma \left( T \right) -\gamma \left( 0 \right) \right]\\
			  &=\text{tr}\left[\int_0^T{\frac{dt}{2\pi}}\partial_t\gamma \right]=C.
\end{split}
\end{equation} 
Furthermore, the Chern number $C$ can be expressed through the non-Abelian Berry curvature $F_{\theta t}$ by applying the Stoke's theorem in the second line of the above equation, i.e.,
\begin{equation}
C=\text{tr}\left[\int_0^T{\frac{dt}{2\pi}}\partial_t\gamma\right]=\int_{0}^{2\pi}d\theta\int_0^Tdt~\text{tr}\left[F_{\theta t}\right],
\end{equation}
where $F_{\theta t}=\partial _{\theta}A_t-\partial _tA_{\theta}+i\left[A_t,A_{\theta}\right] $ with $A_{t}$ and $A_{\theta}$ being the non-Abelian Berry connections defined below Eq.~\eqref{Zak}. The above equation describes an integral over a two-dimensional torus spanned by $\left( \theta ,t \right) $. In numeric, the corresponding non-Abelian Chern number reads~\cite{TFukui2005}
\begin{equation}\label{Chern}
C\equiv \frac{1}{2\pi i}\sum_{l}{F\left( k_{l} \right)},
\end{equation}
where $F\left(k_{l}\right)$ is the field strength. In order to make the numerical process of non-Abelian chern number~\eqref{Chern} readily understandable, we briefly summarize and give a step by step calculation process here, which is essentially the Wilson loop method.

In the first step, we discretized the two-dimensional $\theta$-$t$ plane into $N_{\theta}\times N_t$ grid points. $N_{\theta}$ and $N_t$ denote the total number of grid points on the discrete $\theta$ and $t$ axes, respectively. Each grid point is represented by $k_l=(k_{j_{\theta}},k_{j_t})$, where $k_{j_{\theta}}=2\pi j_{\theta}/N_{\theta}$, $k_{j_t}=Tj_t/N_t$. The corresponding subscripts $l =1,..., N_{\theta} \times N_{t}$, $j_{\theta}=1,...,N_{\theta}$ and $j_t=1,...,N_t$.

In the second step, in order to compute Eq.~\eqref{Chern}, one also need to know the exact expression of $U(2L)$-link variable~\cite{TFukui2005, TFujiwara2001, ML1982}. Let's define the two displacement vectors $\boldsymbol{\hat{\theta}}=\left(2\pi/N_{\theta},0\right)$ and $\boldsymbol{\hat{t}}=\left(0,T/N_t\right)$. And then to compute the Chern number of the degenerate bands one can consider a multiplet $\varphi (k_l)=\left[\left|\varphi_1\left( k_l\right)\right> ,...,\left|\varphi_n\left(k_l\right)\right> ,...,\left|\varphi_{2L}\left(k_l\right)\right> \right]$, where $\left|\varphi_n(k_l)\right> $ denotes the $(L+n)$-th eigenstates of Hamiltonian $H_{\theta}$ of \eqref{hamtheta}. Then, one can get $U\left(2L\right)$-link variables as
\begin{equation}
	\begin{split}
U_{\theta}\left( k_l \right) =\frac{1}{\mathscr{N} _{\theta}\left( k_l \right)}\det \left[ \varphi ^{\dagger}\left( k_l \right) \varphi \left( k_l+\boldsymbol{\hat{\theta}} \right) \right],\\
U_t\left( k_l \right) =\frac{1}{\mathscr{N} _t\left( k_l \right)}\det \left[ \varphi ^{\dagger}\left( k_l \right) \varphi \left( k_l+\boldsymbol{\hat{t}} \right) \right], 
	\end{split}
\end{equation}
with $\mathscr{N}_{\theta}\left(k_l\right)=\left| \det\left[\varphi^{\dagger}\left(k_l\right) \varphi\left(k_l+\boldsymbol{\hat{\theta}}\right) \right]\right|$ and $\mathscr{N}_t\left( k_l\right) =\left|\det\left[\varphi ^{\dagger}\left(k_l \right)\varphi\left(k_l+\boldsymbol{\hat{t}} \right)\right]\right|$ being the normalization constants.

In the third step, one can construct $F\left( k_l \right)$ in terms of $U\left(2L\right)$-link variables~\cite{TFukui2005, TFujiwara2001, ML1982}, i.e.,
\begin{equation}
F\left( k_l \right) =\ln{\left[U_{\theta}\left( k_l \right) U_t\left( k_l+\boldsymbol{\hat{\theta}} \right) U_{\theta}\left( k_l+\boldsymbol{\hat{t}} \right) ^{-1}U_t\left( k_l \right) ^{-1} \right]}.  
\end{equation}
Then, by performing a straightforward calculation, one can obtain the non-Abelian Chern number of Eq.~\eqref{Chern}. We have provided the code for calculating non-Abelian Chern number in the Supplemental Material~\cite{SM}.

To show the competition between the non-Abelian topology and the quasiperiodic disorder, one should first confirm the robustness of the topological pumping. In Fig.~\ref{p1}(b), one can see that the parameter loop makes a non-Abelian topological pumping, which reduces to order case for $W=0$~\cite{VBrosco2021}. As shown in Fig.~\ref{p1}(c), we numerically compute the Chern number as a function of disorder strength $W$. One can find that the quantized Chern number $C=1$ preserves from the clean limit to weak disorder regime. However, for strong disorder, the non-Abelian topological pumping breaks down. This property is similar to the robustness of Abelian Thouless pumping under quasiperiodic disorders or random disorders~\cite{YPWu2022}.

To further discuss the robustness of non-Abelian topological pumping under different disorders, we numerically simulate the corresponding process of adiabatic evolution. Experimentally, one can consider a excitation in photonic waveguide array~\cite{VBrosco2021,YKSun2022}. By modulating their refraction index and relative parameters, the non-Abelian quasiperiodic Hamiltonian and the evolution process can be realized. In numerical simulation, we take $\left|\psi \left( 0 \right) \right>$ as the initial state, which is a localized Wannier state of the degenerate band. This is the excitation at the center of the waveguide arrays. Over a full pumping loop, one can measure the center-of-mass (COM) shift, i.e., $\Delta X_{com}=X_{com}\left(T\right)-X_{com}\left(0\right)$, where $X_{com}\left(t\right)=\sum_j{j\left|\psi _j\right|^2}$ with $\left|\psi _j\right|^2$ as the density at site $j$. 

The values of COM $X_{com}\left(t\right)$ are integer multiple of the unit cell length ($d=2a$). In addition to non-Abelian Chern number, COM shift can also reflect the topological properties of the system~\cite{VBrosco2021}. In Fig.~\ref{p1}(d), we plot the COM over three loop periods for $W=0$, $0.5$, $4.5$, respectively. The results reveal that for $W=0$ (order case) and $0.5$ (weak disorder) the COM shift after the pumping loop $\Delta X_{com}/d\approx~3$, while for $W=4.5$ (strong disorder), $\Delta X_{com}/d\approx~0$. The result demonstrates the robustness of the non-Abelian topological pumping with nearly quantized COM shift for weak disorders.

\begin{figure*}[bhtp] \centering
	\includegraphics[width=17cm]{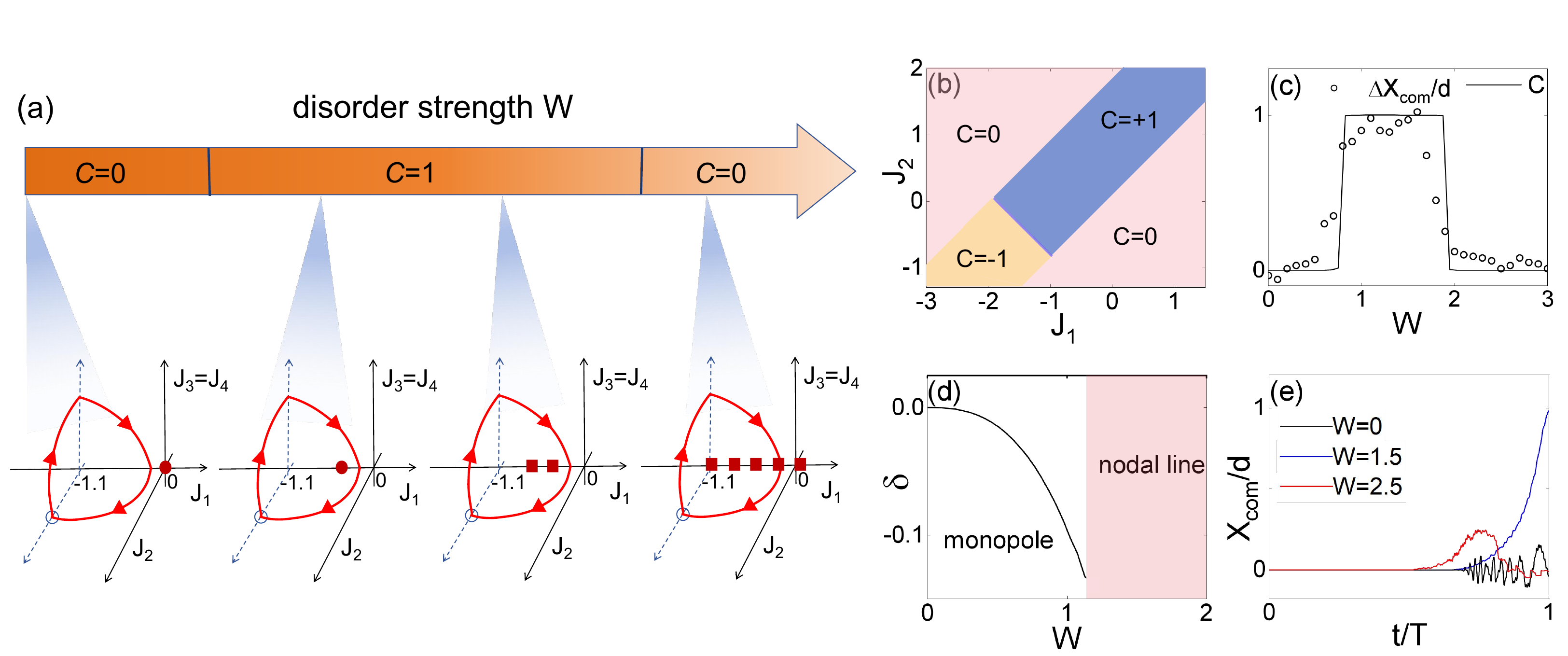}
	\caption{(a) A scheme to manipulate non-Abelian topological pumping by quasiperiodic disorders. Red points indicate the monopole's position and the dashed line means monopole becomes a nodal line. Four plots correspond to $W=0$, $1.1$, $1.5$, $2.5$, respectively. Blue circles present the starting points. (b) Phase diagram in $J_1-J_2$ plane. The corresponding Chern number are marked. (c) The plot of Chern number $C$ with $L=10$ (solid line) and the COM shift $\Delta X_{com}/d$ with ($L=251$)(hollow circle) versus $W$. (d) The monopole position parameter $\delta$ as a function of $W$ for $W<W_l\approx 1.2$. When $W>W_l$ the monopole becomes a nodal line. (e) One loop time evolutions of $X_{com}\left(t\right)/d$ for $W=0$, $1.5$, $2.5$. The COM shift data in (c) and (e) are averaged over $50$ samples with $\beta$ valued from $0\sim 2\pi$ and $L=251$. The four hopping parameters are evolving as shown in Figure (a).} \label{p2}
\end{figure*}

\section{disorder induced non-Abelian topological phase transition}
\label{s3}
Now we discuss how to manipulate the non-Abelian topological phase transition by tunning quasiperiodic disorders. First, we define the monopole as the singularity of the non-Abelian Berry curvature $F_{\theta t}$ in the $(\theta,t)$ space at typical parameters. For the Hamiltonian~(\ref{hamtheta}), the monopole is located at $\left( J_1,J_2,J_3,J_4 \right) =\left( \delta ,0,0,0 \right) $. The $\delta$ can be obtained by solving the following self-consistent equation \eqref{selfconsistent}. When $W=0$, we have $\delta=0$, which is the original point in the Fig.~\ref{p1}(b). To have a pumping with $C=0$, we translate the parameter loop of Fig.~\ref{p1}(b) along the $J_{1}$ axis by $-1.1$ [see Fig.~\ref{p2}(a)]. The corresponding analytic expressions for the new hopping terms have been shown in Appendix~\ref{AEHT}. After the shift of the parameter loop the monopole is out of the topological areas and then the system reflects a topological trivial pumping with $C=0$ in the clean limit~[see Figs.~\ref{p2}(a)(b)]. A quantized non-Abelian topological pumping requires the monopole to stay in the topology non-trivial area of Fig.~\ref{p2}(b). It is then natural to wonder whether quasiperiodic disorder can be used to bring the monopole back into the topological region and thus induces topological pumping. The relative results are shown in Fig.~\ref{p2}.

By increasing the disorder strength $W$, one can find that the monopole moves along the $J_1$ direction and enters the topology non-trivial area, giving rise to a non-Abelian topological pumping with $C=1$. If one continues to increase the disorder strength, the monopole becomes a nodal line, which is similar to the Abelian version~\cite{YPWu2022}. When the nodal line crosses the boundary of the topological area, the non-Abelian topological pumping breaks down and the corresponding $C=0$. The process of emergent non-Abelian topological pumping has been shown in Fig.~\ref{p2}(a). Furthermore, we numerically calculate the corresponding Chern number as the function of the disorder strength $W$ [see Fig.~\ref{p2}(c)]. The results illustrate that weak $W$ can induce the non-Abelian topological pumping. However, such a disorder-induced non-Abelian topological pumping is different from the Anderson Thouless pumping where most of the adiabatic eigenstates are localized. Later, we will show the corresponding IPR to prove that during the process of a non-Abelian topological pumping localized and delocalized eigenstates can coexist.
	
To analyze the mechanism of the emergence of the disorder-induced non-Abelian Thouless pumping, we calculate the disorder-induced shift of the monopole using the self-consistent Born approximation~\cite{YPWu2022}. For weak and moderate disorder, it can be considered as the self energy term $\Sigma\left(W\right)$ to renormalize the Hamiltonian under the clean limit. Then, one can get the self-consistent equation as
\begin{equation}\label{selfconsistent}
\frac{1}{E_f-\mathcal{H}_0\left(k\right)-\Sigma\left(\mathrm{W}\right)}=\left<\frac{1}{E_f-\mathcal{H} _{\mathrm{eff}}\left( \mathrm{k},\mathrm{W} \right)} \right>_{\mathrm{q}},
\end{equation}
where $E_f=0$ is the Fermi energy, and 
\begin{equation}\label{eq:expand_O}
  \mathcal{H} _0\left( k \right) =\left( \begin{matrix}
			0&		J_1+J_2\mathrm{e}^{ik}&		J_3&		J_4\\
			J_1+J_2\mathrm{e}^{-ik}&		0&		0&		0\\
			J_3&		0&		0&		0\\
			J_4&		0&		0&		0\\
	 \end{matrix} \right) ,
\end{equation}
is the Bloch Hamiltonian in the clean limit. $\mathcal{H} _{\mathrm{eff}}$ denotes the effective Hamiltonian renormalized by the quasiperiodic disorder with index $q=1,2,...,N_q$ and $\left< \cdots \right> _{\mathrm{q}}$ denotes averaging over $N_{\mathrm{q}}$ samples. $\mathcal{H} _{\mathrm{eff}}$ can be written as
\begin{equation}\label{Effm}
		\mathcal{H} _{\mathrm{eff}}=\left( \begin{matrix}
			0&		J+J_{2}\mathrm{e}^{ik}&		J_3&		J_4\\
			J+J_{2}\mathrm{e}^{-ik}&		0&		0&		0\\
			J_3&		0&		0&		0\\
			J_4&		0&		0&		0\\
		\end{matrix} \right) ,
\end{equation}
where $J=J_{1}+W\cos \left( 2\pi \alpha q \right)$. By numerically solving the Eq.~\eqref{Effm}, one can obtain  the corresponding $\Sigma\left(\mathrm{W}\right)$ matrix. Since the key shift is on $J_1$, the $(1,2)$ and $(2,1)$ elements of the matrix is the principle term. Then, one can get the position of the monopole by the relation $\delta=-\left[\Sigma\left(W\right)\right]_{21}$, where $\left[\Sigma\left(W\right)\right] _{21}$ denotes the $(2,1)$ element of the matrix [see Fig.~\ref{p2}(d)]. This proves that the monopole can be gradually pulled in the topological region to finally become a nodal line with an ever-increasing disorder.

Then, we show that the disorder-induced non-Abelian topological pumping can be observed in optical systems. Similar to what we have done in Sec.~\ref{s2}, we set the localized Wannier state of the degenerate flat band as the initial state and again calculate the evolution and corresponding COM shift $\Delta X_{com}$. One can find that the disorder-induced non-Abelian Thouless pumping can be observed from the time evolution of the COM as shown in Fig.~\ref{p2}(e). For a topological trivial case [see the $W=0$ line in Fig.~\ref{p2}(e)], the time evolution of the COM is around the origin which demonstrates that $\Delta X_{com}/d\approx 0$. Furthermore, one can find that $\Delta X_{com}/d\approx 1$ for $W=1.5$, which is nearly quantized and similar to the non-Abelian Thouless pumping without disorders in Fig.~\ref{p1}(d). Continue to increase the disorder strength, one can find the breakdown of the topological pumping and the corresponding COM is again around the origin with $\Delta X_{com}/d\approx 0$ [see the $W=2.5$ line in Fig.~\ref{p2}(e)]. The corresponding $\Delta X_{com}$ as a function of $W$ for the lattice $L=251$ is shown in Fig.~\ref{p2}(c). These phenomena indicate the emergence of the disorder-induced non-Abelian Thouless pumping in the weak disorder strength area. The deviation of the quantization mainly comes from the finite-size effect in our simulations. 
	
Finally, let's explain why such a disorder-induced topological pumping can not be dubbed a topological Anderson Thouless pumping. To study the localization effect, we study the IPR, which reads
\begin{equation}\label{OS} 
  I^{\left( n \right)}=\sum_{j=1}^{4L}{\left| \left< j\mid \psi ^{\left( n \right)} \right> \right|^4},
\end{equation}
where $\left|j\right> $ is a $j$-site localized state and $\left|\psi^{\left(n\right)}\right> $ is $n$-th eigenstate. When the system size is large enough ($4L\rightarrow\infty$), one can differentiate between a localized eigenstate and an extended eigenstate by the value of the IPR. The localization criterion says that a eigenstate is localized for $I^{\left(n\right)}\sim\mathcal{O}\left(1\right) $, while $I^{\left(n\right)}\sim\left(4\mathrm{L}\right)^{-1}$ for an extended eigenstate\cite{SNakajima2021, YPWu2022}.

\begin{figure}[htbp] \centering
\includegraphics[width=9cm]{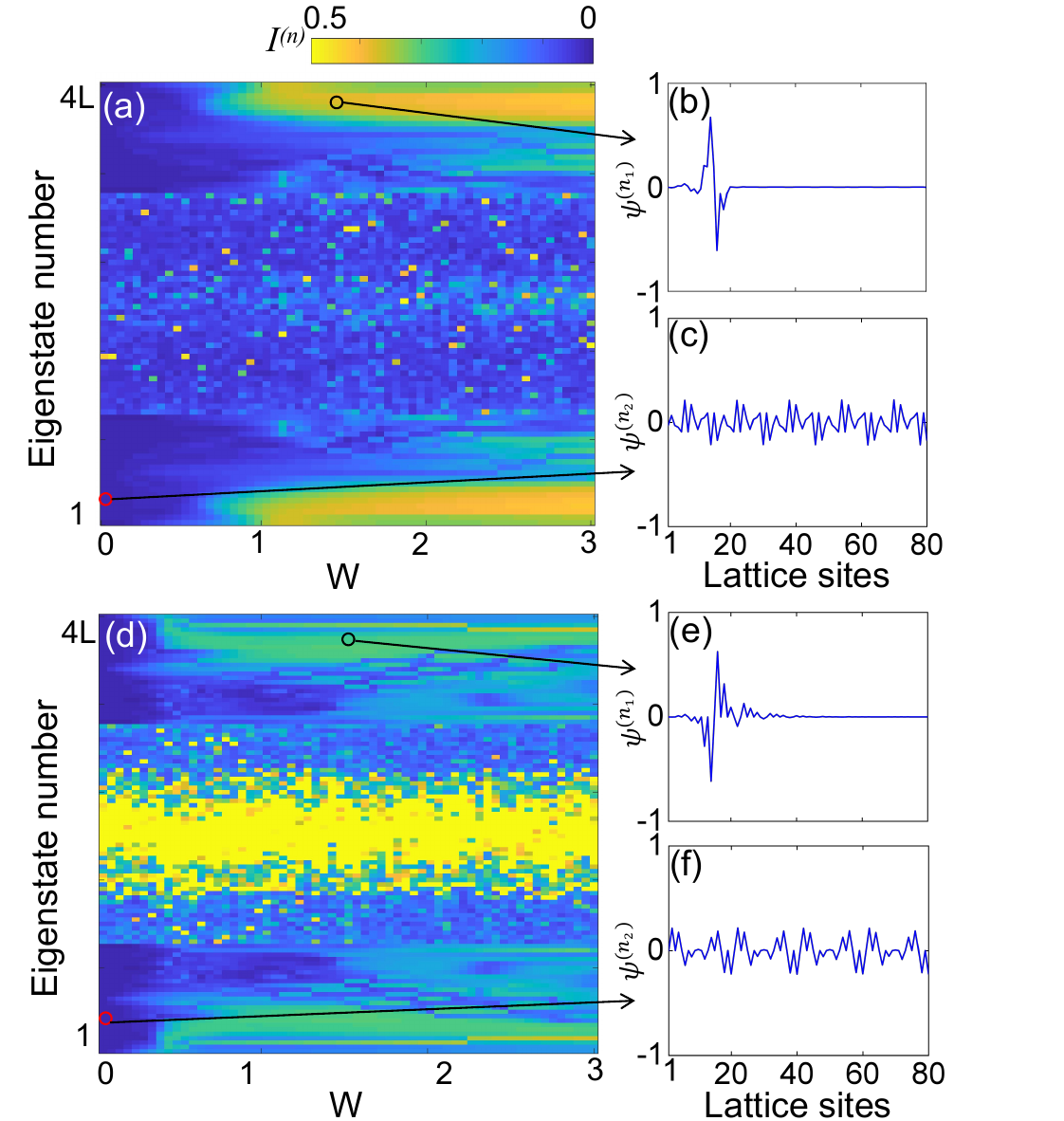}
\caption{(a) The IPR $I^{\left( n \right)}$ as functions of the eigenstate number and the quasiperiodic disorder strength W for $L=20$.  $\left( J_1,J_2,J_3,J_4 \right) =\left(-1.1,\frac{\sqrt{2}}{2},\frac{\sqrt{2}}{2},\frac{\sqrt{2}}{2} \right)$ and $\beta =0$. The right two figures (b) and (c) show the localized and the extended eigenstates indicated by the black and the red circles in (a). $\psi ^{\left(n_1 \right)}$ and $\psi ^{\left( n_2 \right)}$ indicate the $n_1$-th and $n_2$-th eigenstates with $n_1=\frac{15}{4}L, W=1.5$ and $n_2=\frac{L}{4}, W=0$. (d) The IPR $I^{\left( n \right)}$ as functions of the eigenstate number and the quasiperiodic disorder strength W for $L=20$. $\left( J_1,J_2,J_3,J_4 \right) =\left( \frac{\sqrt{2}}{2}-1.1,\frac{\sqrt{2}}{2},0,0 \right)$ and $\beta =0$. Two figures (e) and (f) show the localized and the extended eigenstates indicated by the black and the red circles in (d). $\psi ^{\left(n_1 \right)}$ and $\psi ^{\left( n_2 \right)}$ again indicate the $n_1$-th and $n_2$-th eigenstates with $n_1=\frac{15}{4}L, W=1.5$ and $n_2=\frac{L}{4}, W=0$.}\label{p3}
\end{figure}

We calculate the IPR for a originial $C=0$ loop with an increasing $W$. Without loss of generality, in numerical computation, we choose two points in the parameter loop shown in Fig.~\ref{p2}(a), for the four hopping terms $\left( J_1,J_2,J_3,J_4 \right) =\left(-1.1,\frac{\sqrt{2}}{2},\frac{\sqrt{2}}{2},\frac{\sqrt{2}}{2} \right)$ and $\left( J_1,J_2,J_3,J_4 \right)=\left(\frac{\sqrt{2}}{2}-1.1,\frac{\sqrt{2}}{2},0,0 \right)$, respectively. As shown in Fig.~\ref{p3}(a) and (d), with increasing disorder strength in the zone of flat bands (from $L+1$ to $3L$), the corresponding IPR exhibits the coexistence of localized states and extended states, which origins from the non-Abelian effect. Specifically, this phenomenon would arise in systems where degenerate flat bands exist~\cite{WZhang2023}. In Figs.~\ref{p3}(b)(c), we plot localized and extended eigenstates shown in Fig.~\ref{p3}(a), where the energy level index corresponds to $n_1=\frac{15}{4}L, n_2=\frac{L}{4}$ and the disorder strength for the two eigenstates are $W=1.5, 0$, respectively. In  Figs.~\ref{p3}(e)(f), we also plot localized and extended eigenstates shown in Fig.~\ref{p3}(d) and the energy level index again corresponds to $n_1=\frac{15}{4}L, n_2=\frac{L}{4}$. The disorder strength for the two eigenstates are $W=1.5, 0$, respectively. As the Hamiltonian has degenerate flat bands, the localization properties between the eigenstates of the flat bands and of the non-flat bands are very different. For non-flat bands, the eigenstates change from fully extended states to localized states with an increasing $W$. However, for flat bands, the extended states and the localized states coexist even with an increasing $W$. Furthermore, by comparing Fig.~\ref{p3}(a) and (d), one can find the influence of $J_3$ and $J_4$ on the localization properties. For the case of $J_3=J_4=0$ [see Fig.~\ref{p3}(d)], one can see that the IPR of the middle part (about $1.5L$-$2.5L$) of flat bands is obviously larger than that in Fig.~\ref{p3}(a), which corresponds to more localized eigenstates. The mechanism behind this is: when $J_3$ and $J_4=0$, in a unit cell, the channel connecting sites A and C as well as that connecting sites $A$ and $D$ are closed [see Fig.~\ref{p1}(a)], thus the expansion ability of the corresponding eigenstates of the flat bands is reduced. On the other hand, compared with the presence of AC and AD channels, the closure of $J_3$ and $J_4$ makes hopping between sites A and B easier. Then, the corresponding IPR of $1\sim L$ and $3L+1\sim 4L$ levels in (d) is lower than that in Fig.~\ref{p3}(a). The wave functions of Fig.~\ref{p3}(b)(c) and (e)(f) are in complete agreement with those predicted by IPR.

\section{conclusion}
\label{s4}
In summary, we study non-Abelian Thouless pumping in quasiperiodic disordered systems. Firstly, through theoretical calculation, we show that the non-Abelian Thouless pumping is robust to quasiperiodic disorder. Then we find an emergent non-Abelian Thouless pumping caused by quasiperiodic disorder, and introduce the Chern number of twisted periodic boundary to characterize this phenomenon. The mechanism behind the phenomenon is that the quasiperiodic disorder causes the monopole to move from the topological trivial region into the topological non-trivial region. Furthermore, we numerically simulate such non-Abelian Thouless pumping, and show that the numerical results of centroid evolution are consistent with the theoretical analysis on Chern number. Finally, by calculating the system’s IPR, one can find the non-Abelian inverse Anderson transition is very different from that in the Abelian case, in other words, instead of a clear Anderson transition point, the coexistence of extended state and the localized state emerged. The understanding of non-Abelian topological system with disorder is thus deepened.

\emph{Acknowledgements.}---We thank Dan-Wei Zhang, and Ling-Zhi Tang for useful discussions. This work was supported by the Guangdong Basic and Applied Basic Research Foundation (Grant No.2021A1515012350) and the National Key Research and Development Program of China (Grant No.2022YFA1405300).

	
\begin{appendix}
\section{Analytic expressions for the hopping terms in different time slots}
\label{AEHT}
		
Since the results obtained by introducing disorder to $J_1$, $J_2$ are similar, only the non-Abelian Thouless pumping with disorder by reshaping the terms of $J_1$ ($\rightarrow\tilde{J}_{j}$) is provided in this paper. In the main text, we constructed two kinds of pumping loop as shown in Fig.~\ref{p1}(b) and Fig.~\ref{p2}(a), respectively. 

First, let's discuss the case of Fig.~\ref{p1}(b). Analytic expressions for the hopping terms in Fig.~\ref{p1}(b) can be given as follows,
\begin{equation}
\begin{cases}
				J_1=0
				\\
				J_2=\cos \left( \frac{3\pi}{2T}t \right) 
				\\
				J_3=\sin \left( \frac{3\pi}{2T}t \right) 
				\\
				J_4=J_3
				\\
\end{cases}
\end{equation}
for $t\in\left[0,\frac{T}{3}\right)$, 
\begin{equation}
\begin{cases}
				J_1=\sin \left[ \frac{3\pi}{2T}\left( t-\frac{T}{3} \right) \right] 
				\\
				J_2=0
				\\
				J_3=\cos \left[ \frac{3\pi}{2T}\left( t-\frac{T}{3} \right) \right] 
				\\
				J_4=J_3
				\\
\end{cases}
\end{equation}
for $t\in \left[\frac{T}{3},\frac{2T}{3} \right)$, and
\begin{equation}
\begin{cases}
				J_1=\cos \left[ \frac{3\pi}{2T}\left( t-\frac{2T}{3} \right) \right] 
				\\
				J_2=\sin \left[ \frac{3\pi}{2T}\left( t-\frac{2T}{3} \right) \right] 
				\\
				J_3=0
				\\
				J_4=J_3
				\\
\end{cases}
\end{equation}
for $t\in \left[\frac{2T}{3},T\right]$. Under such circumstance, $W=0$ and $W=0.5$, the corresponding monopole falls into the topological region, so it exhibits quantized topological pumping. However, for $W=4.5$ the nodal line spread across the topological region, so no topological pumping occurs. The results are consistent with the conclusions given by Chern number.

Then, let's turn to discuss the case of Fig.~\ref{p2}(a). The corresponding analytic expression of hopping terms in different time slots reads,
\begin{equation}
\begin{cases}
				J_1=-1.1
				\\
				J_2=\cos \left( \frac{3\pi}{2T}t \right) 
				\\
				J_3=\sin \left( \frac{3\pi}{2T}t \right) 
				\\
				J_4=J_3
				\\
\end{cases}.
\end{equation}
for $t\in \left[0,\frac{T}{3}\right)$,
\begin{equation}
\begin{cases}
				J_1=\sin \left[ \frac{3\pi}{2T}\left( t-\frac{T}{3} \right) \right]-1.1 
				\\
				J_2=0
				\\
				J_3=\cos \left[ \frac{3\pi}{2T}\left( t-\frac{T}{3} \right) \right] 
				\\
				J_4=J_3
				\\
\end{cases}
\end{equation}
for $t\in \left[\frac{T}{3},\frac{2T}{3}\right)$
\begin{equation}
\begin{cases}
				J_1=\cos \left[ \frac{3\pi}{2T}\left( t-\frac{2T}{3} \right) \right]-1.1 
				\\
				J_2=\sin \left[ \frac{3\pi}{2T}\left( t-\frac{2T}{3} \right) \right] 
				\\
				J_3=0
				\\
				J_4=J_3
				\\
\end{cases}
\end{equation}
for $t\in \left[\frac{2T}{3},T\right]$.

This situation is equivalent to placing the monopole outside the topology area at the beginning. Then, one can adjust the value of the disorder ($W$) to bring the monopole back into the topology region. Experimentally, the hopping strength can be easily controlled by varying the distance between waveguides, which enables the control of the non-Abelian Thouless pumping~\cite{YKSun2022}.

\begin{figure}[htbp] \centering
	\includegraphics[width=9cm]{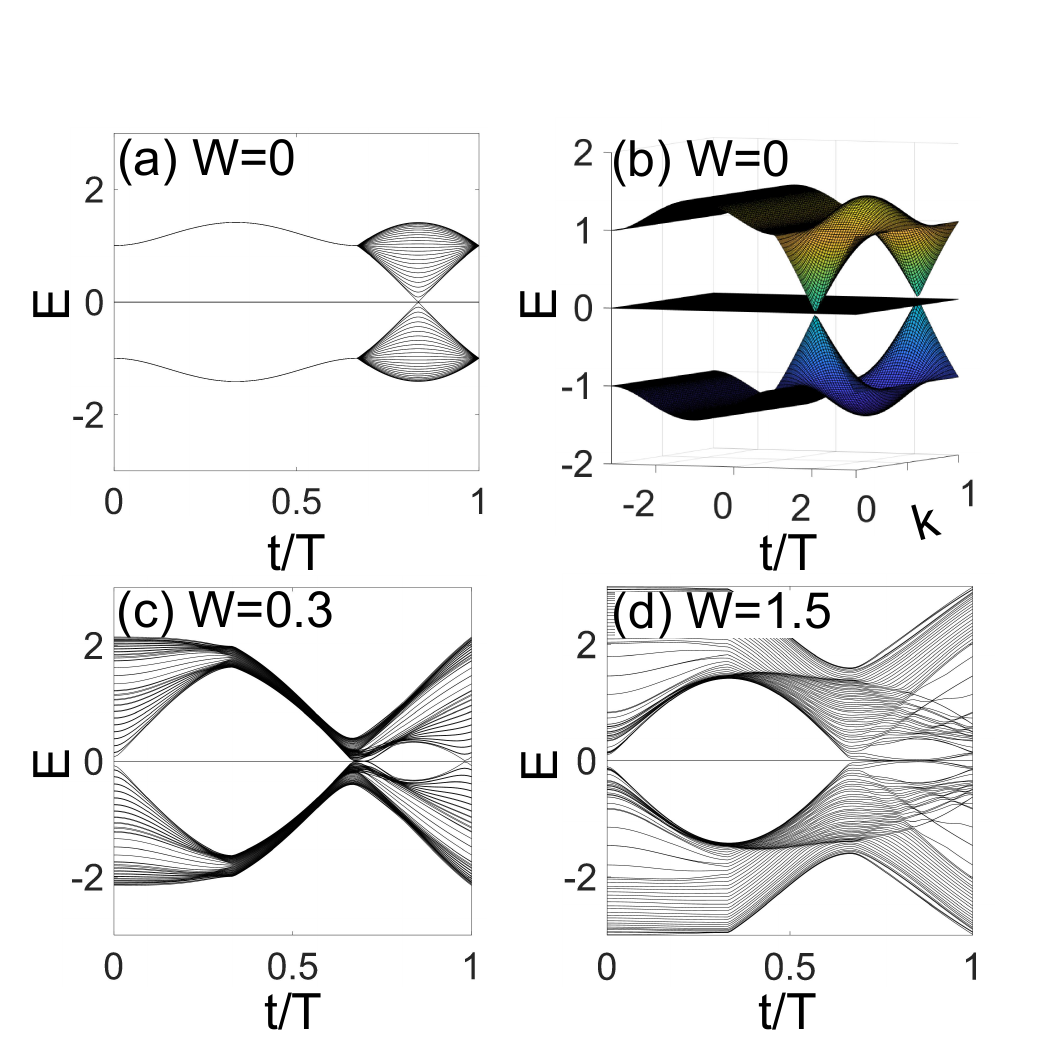}
	\caption{Energy spectrum of Hamiltonian~(\ref{ham}). (a) The plot of the energy spectrum with topological nontrivial phase and without disorder. The four hopping terms are given in Appendix A, the first analytic expressions. (b) The bulk energy spectrum as the function of time $t$ and $k$ without disorder and four hopping terms are the same to (a). (c) The plot of the energy spectrum with topological nontrivial phase and the disorder strength $W=0.3$. The four hopping terms are given in Appendix A, the second analytic expressions. (d) The plot of the energy spectrum with topological trivial phase and the disorder strength $W=1.5$. The four hopping terms are the same to (c). $L=50$, $T=200$ and the data in (c) and (d) are averaged over 50 samples with $\beta$ valued from $0\sim 2\pi $.}\label{p4}
\end{figure}

\section{Energy spectrum of the non-Abelian Hamiltonian}\label{spectrum}
In order to visualize the band structure containing degenerate flat bands, here we plot the energy spectrum of the Hamiltonian in Eq.~\eqref{ham} as shown in Fig.~\ref{p4}. The energy spectrum of non-Abelian Thouless pumping without disorder is shown in Fig.~\ref{p4}(a). The corresponding bulk spectrum is plotted in Fig.~\ref{p4}(b). Both of them tell us that there is degenerate flat band here and no bulk-boundary correspondence. Furthermore, the spectrum with disorder for topolocial trivial case $W=0.3$ (c) and non-trivial case $W=1.5$ (d) are given in Fig.~\ref{p4}(c) and (d), respectively. This shows that even if disorder is introduced, the band structure of the system will not be broken. In other words, the non-Abelian systems discussed in this paper, with or without disorder, have no bulk-boundary correspondence.

\section{Finite size scaling}
\label{FSS}
\begin{figure}[htbp] \centering
	\includegraphics[width=8cm]{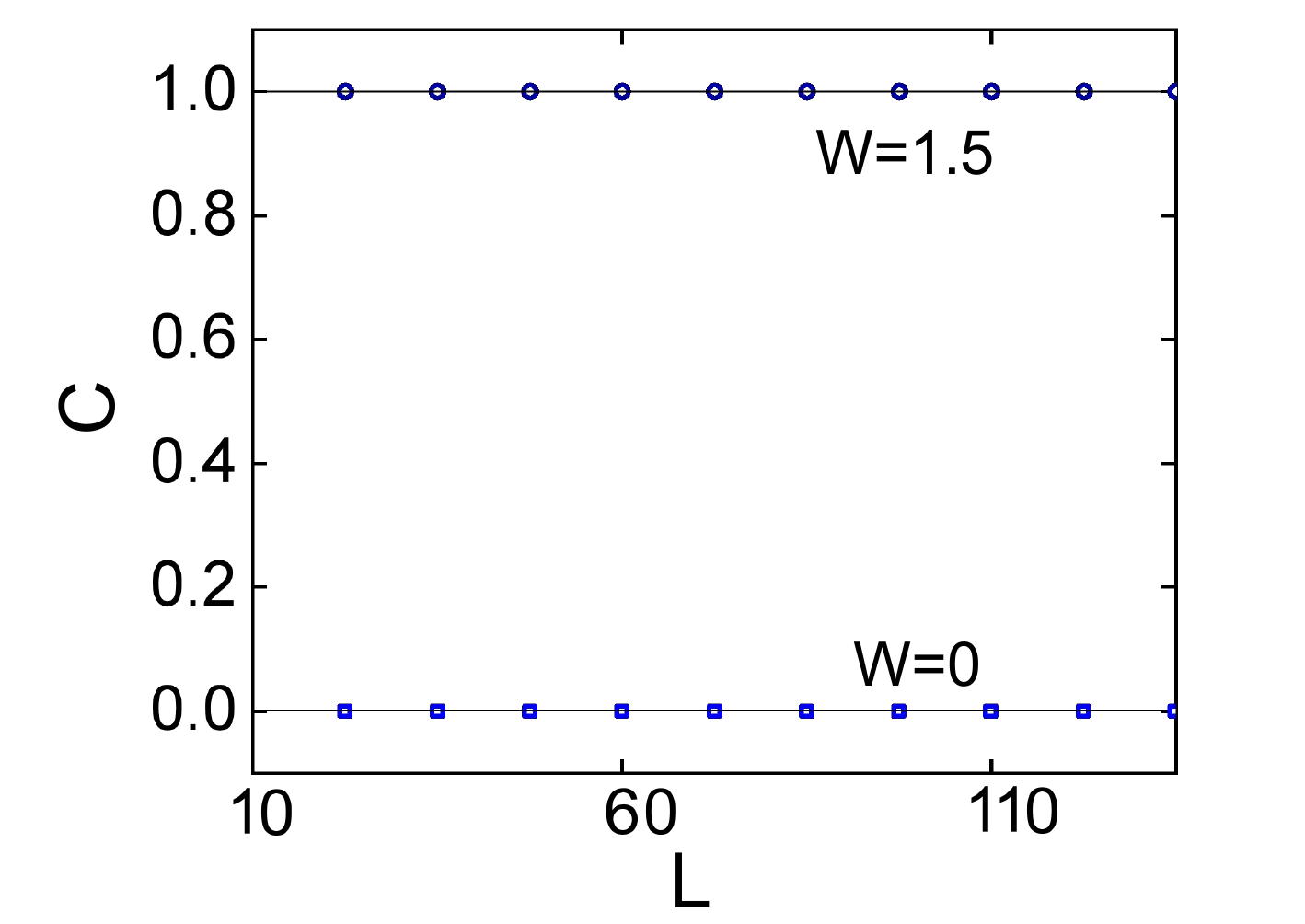}
	\caption{Finite scaling of Chern number $C$ for $W=0, 1.5$. The other parameters are the same as shown in Fig.~\ref{p2}(c) except for $L$. }\label{p5}
\end{figure}

Different from the numerical simulation of the evolution process of COM, the system size does not need to be too large to ensure an accurate non-Abelian Chern number. The condition $L=10$ in main text is enough. Here we conduct a finite-size scaling of Chern number $C$ and plot the Chern number as a function of different $L$ in Fig.~\ref{p5}. In the computation process, we fix $W=0,1.5$, which correspond to $C=0,1$, respectively. One can find that the corresponding Chern numbers preserve for various $L$.

\end{appendix}

\bibliographystyle{apsrev4-2}

\begin{thebibliography}{99}
\bibitem{DJThouless1983} D. J. Thouless, Phys. Rev. B {\bf27}, 6083 (1983).

\bibitem{KvKlitzing1980} K.v.Klitzing, G. Dorda, and M. Pepper, Phys. Rev. Lett. {\bf45}, 494 (1980).

\bibitem{QNiu1984} Q. Niu and D. J. Thouless, J. Phys. A: Math. Gen {\bf17}, 2453 (1984).

\bibitem{QNiu1985} Q. Niu, D. J. Thouless, and Y.-S. Wu, Phys. Rev. B {\bf31}, 3372 (1985).

\bibitem{DXiao2010} D. Xiao, M.-C. Chang, and Q. Niu, Rev. Mod. Phys. {\bf82}, 1959 (2010).

\bibitem{XLQi2011} X.-L. Qi and S.-C. Zhang, Rev. Mod. Phys. {\bf83}, 1057 (2011).

\bibitem{MZHasan2010} M. Z. Hasan and C. L. Kane, Rev. Mod. Phys. {\bf82}, 3045 (2010).

\bibitem{DWZhang2018} D.-W. Zhang, Y.-Q. Zhu, Y. Zhao, H. Yan, and S.-L. Zhu, Adv. Phys. {\bf67}, 253 (2018).

\bibitem{NRCooper2019} N. R. Cooper, J. Dalibard, and I. B. Spielman, Rev. Mod. Phys. {\bf91}, 015005 (2019).

\bibitem{NGoldman2016} N. Goldman, J. C. Budich, and P. Zoller, Nat. Phys. {\bf12}, 639 (2016).

\bibitem{TOzawa2019} T. Ozawa, H. M. Price, A. Amo, N. Goldman, M. Hafezi, L. Lu, M. C. Rechtsman, D. Schuster, J. Simon, O. Zilberberg, and I. Carusotto, Rev. Mod. Phys. {\bf91}, 015006 (2019).

\bibitem{MLohse2016} M. Lohse, C. Schweizer, O. Zilberberg, M. Aidelsburger, and I. Bloch, Nat. Phys. {\bf12}, 350 (2016).

\bibitem{SNakajima2016} S. Nakajima, T. Tomita, S. Taie, T. Ichinose, H. Ozawa, L. Wang, M. Troyer, and Y. Takahashi, Nat. Phys. {\bf12}, 296 (2016).

\bibitem{HILu2016} H.-I. Lu, M. Schemmer, L. M. Aycock, D. Genkina,S. Sugawa, and I. B. Spielman, Phys. Rev. Lett. {\bf116}, 200402 (2016).

\bibitem{CSchweizer2016} C. Schweizer, M. Lohse, R. Citro, and I. Bloch, Phys. Rev. Lett. {\bf117}, 170405 (2016).

\bibitem{SNakajima2021} S. Nakajima, N. Takei, K. Sakuma, Y. Kuno, P. Marra, and Y. Takahashi, Nat. Phys. {\bf17}, 844 (2021).

\bibitem{AFabre2022} A. Fabre, J.-B. Bouhiron, T. Satoor, R. Lopes, and S. Nascimbene, Phys. Rev. Lett. {\bf128}, 173202 (2022).

\bibitem{JMinguzzi2022} J. Minguzzi, Z. Zhu, K. Sandholzer, A.-S. Walter, K. Viebahn, and T. Esslinger, Phys. Rev. Lett. {\bf129}, 053201 (2022).

\bibitem{LWang2013} L. Wang, M. Troyer, and X. Dai, Phys. Rev. Lett. {\bf111}, 026802 (2013).

\bibitem{YEKraus2012} Y. E. Kraus, Y. Lahini, Z. Ringel, M. Verbin, and O. Zilberberg, Phys. Rev. Lett. {\bf109}, 106402 (2012).

\bibitem{ACerjan2020} A. Cerjan, M. Wang, S. Huang, K. P. Chen, and M. C. Rechtsman, Light Sci. Appl. {\bf9}, 178 (2020).

\bibitem{QCheng2022} Q. Cheng, H. Wang, Y. Ke, T. Chen, Y. Yu, Y. S. Kivshar, C. Lee, and Y. Pan, Nat. Commun. {\bf13}, 249 (2022).

\bibitem{WLiu2022} W. Liu, C. Wu, Y. Jia, S. Jia, G. Chen, and F. Chen, Phys. Rev. A {\bf105}, L061502 (2022).

\bibitem{YKe2016} Y. Ke, X. Qin, F. Mei, H. Zhong, Y. S. Kivshar, and C. Lee, Laser Photon. Rev. {\bf10}, 995 (2016).

\bibitem{Wcheng2020} W. Cheng, E. Prodan, and C. Prodan, Phys. Rev. Lett. {\bf125}, 224301 (2020).

\bibitem{IHGrinberg2020} I. H. Grinberg, M. Lin, C. Harris, W. A. Benalcazar, C. W. Peterson, T. L. Hughes, and G. Bahl, Nat. Commun. {\bf11}, 974 (2020).

\bibitem{Mjurgensen2021} M. Jürgensen, S. Mukherjee, and M. C. Rechtsman, Nature {\bf596}, 63 (2021).

\bibitem{Mjurgensen2022} M. Jürgensen, and M. C. Rechtsman, Phys. Rev. Lett. {\bf128}, 113901 (2022).

\bibitem{QFu2022} Q. Fu, P. Wang, Y. V. Kartashov, V. V. Konotop, and F. Ye, Phys. Rev. Lett. {\bf128}, 154101 (2022).

\bibitem{NMostaan2022} N. Mostaan, F. Grusdt, and N. Goldman, Nat. Commun. {\bf13}, 5997 (2022).

\bibitem{MJurgensen2023} M. Jürgensen, S. Mukherjee, C. Jörg, and M. C. Rechtsman, Nat. Phys. {\bf19}, 420 (2023).

\bibitem{RGawatz2022} R. Gawatz, A. C. Balram, E. Berg, N. H. Lindner, and M. S. Rudner, Phys. Rev. B {\bf105}, 195118 (2022).

\bibitem{JAMarks2021} J. A. Marks, M. Schüler, J. C. Budich, and T. P. Devereaux, Phys. Rev. B {\bf103}, 035112 (2021).

\bibitem{YKe2017} Y. Ke, X. Qin, Y. S. Kivshar, and C. Lee, Phys. Rev. A {\bf95}, 063630 (2017).

\bibitem{TSZeng2015} T.-S. Zeng, C. Wang, and H. Zhai, Phys. Rev. Lett. {\bf115}, 095302 (2015).

\bibitem{ASWalter} A.-S. Walter, Z.-J. Zhu, M. Gächter, J. Minguzzi, S. Roschinski, K. Sandholzer, K. Viebahn, and T. Esslinger, Nat. Phys. {\bf19}, 1417 (2023).

\bibitem{YKe2023} Y. Ke and C. Lee, Nat. Phys. {\bf19}, 1387 (2023).

\bibitem{WABenalcazar2022} W. A. Benalcazar, J. Noh, M. Wang, S. Huang, K.-P. Chen, and M. C. Rechtsman, Phys. Rev. B {\bf105}, 195129 (2022).

\bibitem{JFWienand2022} J. F. Wienand, F. Horn, M. Aidelsburger, J. Bibo, and F. Grusdt, Phys. Rev. Lett. {\bf128}, 246602 (2022).

\bibitem{BLWu2022} B.-L. Wu, A.-M. Guo, Z.-Q. Zhang, and H. Jiang, Phys. Rev. B {\bf106}, 165401 (2022).

\bibitem{CYuce2019} C. Yuce, Phys. Rev. A {\bf99}, 032109 (2019).

\bibitem{WHu2017} W. Hu, H. Wang, P. P. Shum, and Y. D. Chong, Phys. Rev. B {\bf95}, 184306 (2017).

\bibitem{ZFedorova2020} Z. Fedorova, H. Qiu, S. Linden, and J. Kroha, Nat. Commun. {\bf11}, 3758 (2020).

\bibitem{RCitro2023} R. Citro and M. Aidelsburger, Nat. Rev. Phys. {\bf5}, 87 (2023).

\bibitem{VBrosco2021} V. Brosco, L. Pilozzi, R. Fazio, and C. Conti, Phys. Rev. A {\bf103}, 063518 (2021).

\bibitem{OYou2022} O. You, S. Liang, B. Xie, W. Gao, W. Ye, J. Zhu, and S. Zhang, Phys. Rev. Lett. {\bf128}, 244302 (2022).

\bibitem{YKSun2022} Y.-K. Sun, X.-L. Zhang, F. Yu, Z.-N. Tian, Q.-D. Chen, and H.-B. Sun, Nat. Phys. {\bf18}, 1080 (2022).

\bibitem{YYang2023} Y. Yang, B. Yang, G. Ma, J. Li, S. Zhang, and C. Chan, arXiv:2305.12206 (2023).

\bibitem{MParto2023} M. Parto, C. Leefmans, J. Williams, F. Nori, and A. Marandi, Nat. Commun. {\bf14}, 1440 (2023).

\bibitem{PHauke2012} P. Hauke, O. Tieleman, A. Celi, C. Ölschläger, J. Simonet, J. Struck, M. Weinberg, P. Windpassinger, K. Sengstock, M. Lewenstein, et al., Phys. Rev. Lett. {\bf109}, 145301 (2012).

\bibitem{LLepori2016} L. Lepori, I. C. Fulga, A. Trombettoni, and M. Burrello, Phys. Rev. A {\bf94}, 053633 (2016).

\bibitem{YYang2019} Y. Yang, C. Peng, D. Zhu, H. Buljan, J. D. Joannopoulos, B. Zhen, and M. Soljačić, Science {\bf365}, 1021 (2019).

\bibitem{QXLv2023} Q.-X. Lv, H.-Z. Liu, Y.-X. Du, L.-Q. Chen, M. Wang, J.-H. Liang, Z.-X. Fu, Z.-Y. Chen, H. Yan, and S.-L. Zhu, arXiv:2305.05849 (2023).

\bibitem{DCheng2023} D. Cheng, K. Wang, and S. Fan, Phys. Rev. Lett. {\bf130}, 083601 (2023).

\bibitem{GPalumbo2021} G. Palumbo, Phys. Rev. Lett. {\bf126}, 246801 (2021).

\bibitem{FWilczek1984} F. Wilczek, and A. Zee, Phys. Rev. Lett. {\bf52}, 2111 (1984).

\bibitem{DWZhang2020} D.-W. Zhang, L.-Z. Tang, L.-J. Lang, H. Yan, and S.-L. Zhu, Sci. China Phys. Mech. Astron. {\bf63}, 267062 (2020).

\bibitem{SLZhuHFu2006} S.-L. Zhu, H. Fu, C.-J. Wu, S.-C. Zhang, and L.-M. Duan, Phys. Rev. Lett. {\bf97}, 240401 (2006).

\bibitem{LBShao2008} L.-B. Shao, S.-L. Zhu, D. Xing, and Z. Wang, Phys. Rev. Lett. {\bf101}, 246810 (2008).

\bibitem{SLZhu2007} S.-L. Zhu, B. Wang, and L.-M. Duan, Phys. Rev. Lett. {\bf98}, 260402 (2007).

\bibitem{SLZhu2006} S.-L. Zhu, Phys. Rev. Lett. {\bf96}, 077206 (2006).

\bibitem{XShen2018} X. Shen and Z. Li, Phys. Rev. A {\bf97}, 013608 (2018).

\bibitem{XShen2019} X. Shen, F. Wang, Z. Li, and Z. Wu, Phys. Rev. A {\bf100}, 062514 (2019).

\bibitem{XDHu2021} X.-D. Hu, L.-Y. Li, Z.-X. Guo, and Z. Li, New. J. Phys. {\bf23}, 073031 (2021).

\bibitem{ZXGuo2022} Z.-X. Guo, X.-J. Yu, X.-D. Hu, and Z. Li, Phys. Rev. A {\bf105}, 053311 (2022).

\bibitem{XShen2022} X. Shen, Y.-Q. Zhu, and Z. Li, Phys. Rev. B {\bf106}, L180301 (2022).

\bibitem{HTDing2023} H.-T. Ding, C.-X. Zhang, J.-X. Liu, J.-T. Wang, D.-W. Zhang, and S.-L. Zhu, arXiv:2312.01086.

\bibitem{JQin} J. Qin and H. Guo, Phys. Lett. A {\bf380}, 2317 (2016).

\bibitem{MMWauters2019} M. M. Wauters, A. Russomanno, R. Citro, G. E. Santoro, and L. Privitera, Phys. Rev. Lett. {\bf123}, 266601 (2019).

\bibitem{ALCHayward2021} A. L. C. Hayward, E. Bertok, U. Schneider, and F. H. Meisner, Phys. Rev. A {\bf103}, 043310 (2021).

\bibitem{JLi2009} J. Li, R.-L. Chu, J. K. Jain, and S.-Q. Shen, Phys. Rev. Lett. {\bf102}, 136806 (2009).

\bibitem{CWGroth2009} C. W. Groth, M. Wimmer, A. R. Akhmerov, J. Tworzydło, and C. W. J. Beenakker, Phys. Rev. Lett. {\bf103}, 196805 (2009).


\bibitem{CZChen2015} C.-Z. Chen, J. Song, H. Jiang, Q.-F Sun, Z. Wang, and X.-C. Xie, Phys. Rev. Lett. {\bf115}, 246603 (2015).

\bibitem{HMGuo2010} H.-M. Guo, G. Rosenberg, G. Refael, and M. Franz, Phys. Rev. Lett. {\bf105}, 216601 (2010).

\bibitem{AAltland2014} A. Altland, D. Bagrets, L. Fritz, A. Kamenev, and H. Schmiedt, Phys. Rev. Lett. {\bf112}, 206602 (2014).

\bibitem{EJMeier} E. J. Meier, F. A. An, A. Dauphin, M. Maffei, P. Massignan, T. L. Hughes, and B. Gadway, Science {\bf362}, 929
(2018).

\bibitem{SStutzer2018} S. Stützer, Y. Plotnik, Y. Lumer, P. Titum, N. H. Lindner, M. Segev, M. C. Rechtsman, and A. Szameit, Nature {\bf560}, 461 (2018).

\bibitem{GGLiu2020} G.-G. Liu, Y. Yang, X. Ren, H. Xue, X. Lin, Y.-H. Hu, H.-x. Sun, B. Peng, P. Zhou, Y. Chong, and B. Zhang, Phys. Rev. Lett. {\bf125}, 133603 (2020).

\bibitem{LZTang2022} L.-Z. Tang, S.-N. Liu, G.-Q. Zhang, and D.-W. Zhang, Phys. Rev. A {\bf105}, 063327 (2022).

\bibitem{XHCui2022} X. Cui, R.-Y. Zhang, Z.-Q. Zhang, and C. T. Chan, Phys. Rev. Lett. {\bf129}, 043902 (2022).

\bibitem{LZTang2020} L.-Z. Tang, L.-F. Zhang, G.-Q. Zhang, and D.-W. Zhang, Phys. Rev. A {\bf101}, 063612 (2020).

\bibitem{HCHsu2020} H.-C. Hsu and T.-W. Chen,
Phys. Rev. B {\bf102}, 205425 (2020).

\bibitem{YPWu2022} Y.-P. Wu, L.-Z. Tang, G.-Q. Zhang, and D.-W. Zhang, Phys. Rev. A {\bf106}, L051301 (2022).

\bibitem{WZhang2023} W. Zhang, H. Wang, H. Sun, and X. Zhang, Phys. Rev. Lett. {\bf130}, 206401 (2023).

\bibitem{llin2023} L. Lin, Y.-G. Ke, L. Zhang, and C.-h. Lee, Phys. Rev. B {\bf108}, 174204 (2023).

\bibitem{Bohm2003} A. Bohm, A. Mostafazadeh, H. Koizumi, Q. Niu, and J. Zwanziger, 2003, The Geometric Phase in Quantum Systems: Foundations, Mathematical Concepts, and Applications in Molecular and Condensed Matter Physics, Springer-Verlag, Berlin.

\bibitem{TFukui2005} T. Fukui, Y. Hatsugai, and H. Suzuki, J. Phys. Soc. Jpn. {\bf74}, 1674
(2005).

\bibitem{TFujiwara2001} T. Fujiwara, H. Suzuki, and K. Wu, Prog. Theor. Phys. {\bf105}, 789
(2001).

\bibitem{ML1982} M. Lüscher, Commun. Math. Phys.  {\bf85}, 39 (1982).

\bibitem{SM} See Supplemental Material for Python code.
\end{thebibliography}


\end{document}